\documentclass[conference]{IEEEtran}
\IEEEoverridecommandlockouts
\usepackage{cite}
\usepackage[shortlabels]{enumitem}
\usepackage{braket}
\usepackage{amsmath,amssymb,amsfonts}
\usepackage{epsfig,latexsym,graphicx}
\usepackage{epstopdf}
\usepackage{algorithmic}
\usepackage{graphicx}
\usepackage{textcomp}
\usepackage{xcolor}
\usepackage{bbm}
\usepackage[T1]{fontenc}
\usepackage[utf8]{inputenc}
\usepackage[english]{babel}
\usepackage{multirow}
\usepackage{blkarray}
\usepackage{subcaption}
\usepackage{tablefootnote}
\usepackage{tabularx}

\usepackage[font=small,labelfont=bf]{caption}
\newtheorem{theorem}{\bf Theorem}

\newtheorem{definition}{\bf Definition}
\newtheorem{remark}{\bf Remark}

\DeclareMathOperator*{\argmax}{argmax}

\begin{document}
\title{On the Capacity Region of a Quantum Switch with Entanglement Purification}
\author{
\IEEEauthorblockN{Nitish K. Panigrahy\IEEEauthorrefmark{1},
Thirupathaiah Vasantam \IEEEauthorrefmark{2}, Don Towsley \IEEEauthorrefmark{3} and Leandros Tassiulas\IEEEauthorrefmark{1}}

\IEEEauthorblockA{\IEEEauthorrefmark{1}Yale University, USA. \IEEEauthorrefmark{2}Durham University, UK. 
\IEEEauthorrefmark{3}University of Massachusetts Amherst, USA.  \\
Email: \IEEEauthorrefmark{1}\{nitishkumar.panigrahy, leandros.tassiulas\}@yale.edu, \IEEEauthorrefmark{2}thirupathaiah.vasantam@durham.ac.uk, \IEEEauthorrefmark{3}towsley@cs.umass.edu}
}
\maketitle
\thispagestyle{plain}
\pagestyle{plain}

\begin{abstract}
Quantum switches are envisioned to be an integral component of future entanglement distribution networks. They can provide high quality entanglement distribution service to end-users by performing quantum operations such as entanglement swapping and entanglement purification. In this work, we characterize the capacity region of such a quantum switch under noisy channel transmissions and imperfect quantum operations. We express the capacity region as a function of the channel and network parameters (link and entanglement swap success probability), entanglement purification yield and application level parameters (target fidelity threshold). In particular, we provide necessary conditions to verify if a set of request rates belong to the capacity region of the switch. We use these conditions to find the maximum achievable end-to-end user entanglement generation throughput by solving a set of linear optimization problems. We develop a max-weight scheduling policy and prove that the policy stabilizes the switch for all feasible request arrival rates. As we develop scheduling policies, we also generate new results for computing the conditional yield distribution of different classes of purification protocols. From numerical experiments, we discover that performing link-level entanglement purification followed by entanglement swaps provides a larger capacity region than doing entanglement swaps followed by end-to-end entanglement purification. The conclusions obtained in this work can yield useful guidelines for subsequent quantum switch designs.
\end{abstract}

\begin{IEEEkeywords}
Quantum Switch; Capacity Region; Entanglement Purification; Max-weight Scheduling
\end{IEEEkeywords}

\section{Introduction}
Entanglement distribution is critical for many promising quantum applications such as quantum key distribution (QKD) \cite{Shor00}, teleportation \cite{Ma12}, quantum sensing \cite{Eldredge18}, clock
synchronisation \cite{komar2014quantum}, and distributed quantum computing \cite{Broadbent09}. Recent works \cite{zhao2022e2e,pant2019routing,cicconetti2021request,chakraborty2019distributed} and proposals \cite{lloyd2004infrastructure,dahlberg2019link}, have given much attention to distribute entanglements in a large scale quantum network that supports these applications. In an entanglement distribution network, a network of quantum switches are connected through optical fiber links. The switches are responsible for creating entanglements with their neighbors and perform a quantum operation known as {\it entanglement swapping} on individual entanglements to create {\it end-to-end user entanglements}.

\begin{figure}[!htbp]
\centering
\begin{minipage}{0.35\textwidth}
\includegraphics[width=1\textwidth]{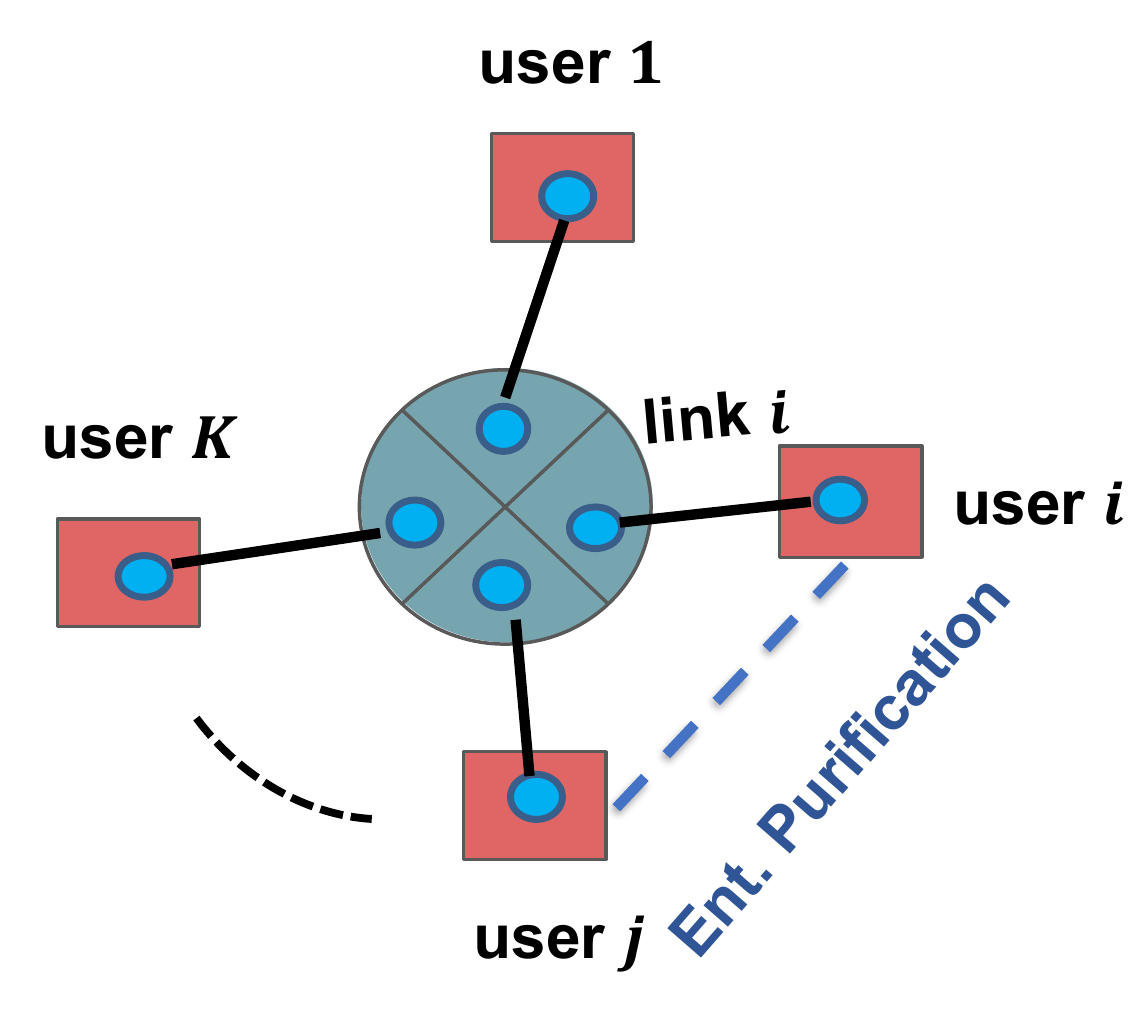}
\subcaption{}
\end{minipage}
\caption{A quantum switch serving end-to-end entanglement to $K$ end users.}
\label{switch}
\end{figure}

In this work, we focus on a star shaped quantum network represented by a quantum switch and a set of $K$ end user nodes as shown in Figure \ref{switch}. End users are connected to the switch via a quantum communication channel. An entanglement generation source located in the middle of the channel creates maximally entangled bipartite Bell states ({\it EPR pairs}) and sends one half of the pair to the switch and the other to the end user. 

This shared entangled state between the switch and an end user is known as {\it link-level entanglement}. The switch and each end user can generate multiple such link-level entanglements. Requests to create end-to-end entanglement between users arrive at the switch. The role of the switch is to select specific link level entanglements to consume and perform entanglement swapping thereby generating end-to-end user entanglements to satisfy the requests. The quantum applications at the end users can then consume these end-to-end entanglements or keep them at their respective quantum memories for future usage.

Recently, there has been significant interest in analyzing the performance of a single quantum switch that serves multiple end users  \cite{Nain20,Vardoyan21,Vardoyan20,Vasantam22, Dai21}. For example, Vardoyan et al. \cite{Vardoyan21,Vardoyan20} derived expressions for the maximum possible end-to-end bipartite entanglement generation rate of a quantum switch using discrete-time and continuous-time Markov chains. Nain et al. \cite{Nain20} derived closed form expressions for the capacity of a switch serving multi-partite entanglements to end users. They also derived the necessary and sufficient conditions for stability of the switch. However, these works \cite{Nain20,Vardoyan21,Vardoyan20} assume the creation of end-to-end entanglements in a continuous manner, i.e., once the link-level entanglements are available, they are immediately consumed to create end-to-end entanglements. The notion of serving entanglement requests from quantum applications in an on-demand manner and scheduling those requests was not considered in these works. 

The request scheduling problem in a quantum switch primarily involves determining which link-level entanglements should be consumed during entanglement swapping and at what times so that respective end user entanglement requests can be satisfied. The scheduling problem can further have multiple scheduling objectives such as maximizing the aggregate entanglement generation rate or maintaining finite request backlogs there by providing finite end-user latencies. The latter objective is also known as achieving stability for the switch and has been investigated in \cite{Vasantam22, Dai21} under different assumptions on lifetimes of link-level and end-to-end entanglements. In particular, Vasantam et al. \cite{Vasantam22} assumed entanglements to last for a single time slot and characterized the {\it capacity region} of the switch. Here, the capacity region describe the set of end-to-end request rates that the switch can stably support with finite request backlogs. Similarly, Dai et al. \cite{Dai21} considered infinite entanglement lifetime and proposed scheduling protocols that stabilize the switch.

While the aforementioned works on both continuous and request based models consider channel loss and entanglement swap failures, they assume the channel and quantum operations to be noise-free and do not explicitly model imperfections in capacity region calculations. Quantum transmissions are inherently noisy due to their interactions with the environment and generally produce non-maximally entangled link-level and end-to-end entanglements. Thus, the generated entanglements may not be good enough for consumption at the application level. In this work, we remove the noise-free assumption and study the problem of quantum switch scheduling with noisy channels and imperfect quantum operations. We associate a target entanglement fidelity threshold ($F^{th}$) for quantum applications where fidelity is a widely used metric to quantify the quality of an entanglement. We consider the request for entanglement generation to be satisfied only when the generated end-to-end entanglement has a fidelity greater than $F^{th}.$

Due to noise in quantum channels and devices, fidelity of an entangled state decreases with each quantum operation. Let $F_{link}$ and $F_{swap}$ denote the fidelity of the link-level entanglement and the entangled state created after a swap respectively. Typically, we have $F_{swap} < F_{link}  < 1$, where the fidelity of a perfect entangled state is considered to be $1$. When $F_{swap} < F^{th}$, the generated user entanglements are not good enough for consumption at the application level. To circumvent this issue, one can perform nested entanglement purification \cite{Dur1999} on these low quality entanglements to generate fewer number of high quality entanglements that achieve a target fidelity. 

The order in which entanglement purification and entanglement swap operations are performed can lead to the following two architectures. (i) Purify $\&$ Swap (PS): One where purification is performed on link-level entanglements, followed by entanglement swaps. (ii) Swap $\&$ Purify (SP): One that performs entanglement swapping first on link-level entanglements and then performs purification on end-to-end entanglements. It is important to understand the trade-offs and design constraints of these architectures, which can shed some light on the future design of quantum switches. To this end, we characterize and compare the capacity regions of a quantum switch under both of these architectures.

\subsection{Contributions}
We characterize the capacity region of a quantum entanglement distribution switch under channel noise and imperfect quantum operations. Our contributions are summarized below.
\begin{itemize}
\item We determine the capacity region of a quantum switch as a function of the network (link $\&$ swap entanglement success probability, $F_{link}$), entanglement purification (conditional yield) and application level ($F^{th}$) parameters under PS and SP architectures.
\item We develop max-weight scheduling policies for both PS and SP architectures that stabilize the switch for all feasible request arrival rates.
\item We generate new results for computing the conditional yield distribution of different classes of purification protocols.
\item We evaluate and compare the capacity regions of PS and SP architectures under different classes of purification protocols, channel and network parameters.
\end{itemize}
The rest of the paper is organized as follows. In Section \ref{sec:prelim}, we briefly introduce the quantum background and the system model relevant to this work. Section \ref{capacity} discusses some of the main results related to determining the capacity region of a switch under PS and SP architecture. Section \ref{expected_yield} presents results related to deriving the conditional yield distribution and expected conditional yield of different classes of entanglement purification protocols. We evaluate the performance of purification protocols and PS, SP architectures in Section \ref{performance}. Section \ref{conclusion} concludes the paper and discusses some possible future work.

\section{Technical Preliminaries}\label{sec:prelim}
\begin{figure*}[htbp]
\centering
\begin{minipage}{0.45\textwidth}
\includegraphics[width=1\textwidth]{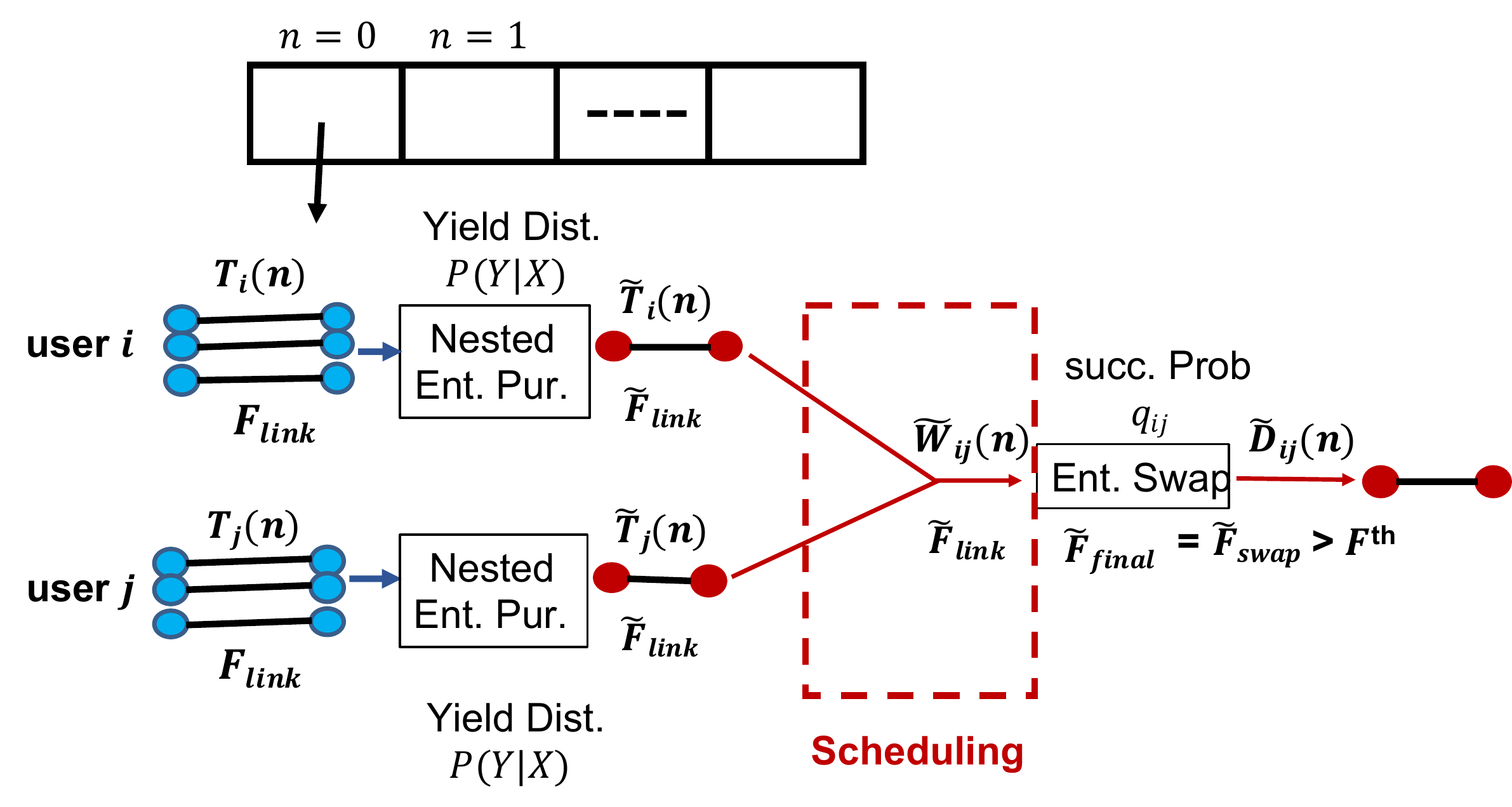}
\subcaption{}
\end{minipage}
\hspace{1cm}
\begin{minipage}{0.45\textwidth}
\includegraphics[width=1\textwidth]{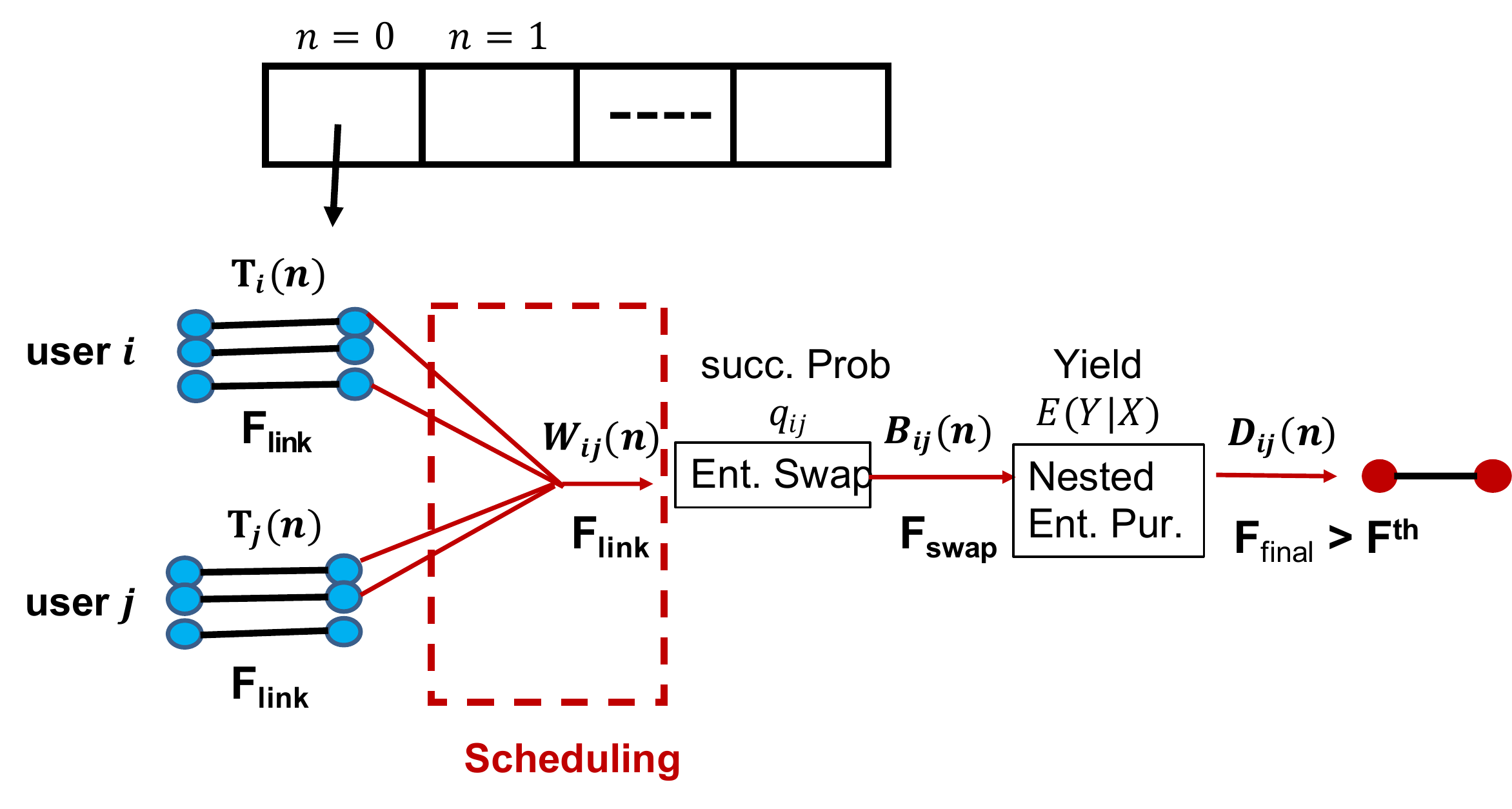}
\subcaption{}
\end{minipage}\hfill
\caption{(a) Purify and Swap architecture (b) Swap and Purify architecture in time slot $n$.}
\label{stability}
\end{figure*}

We consider a quantum switch that creates entanglements between $K$ end users as shown in Figure \ref{switch}. User $i$ is connected to the switch via a fiber optic link. For brevity, we will use the term ``node" to refer to both end users and the quantum switch with node $0$ referring to the switch. 
\subsection{Quantum Background} 
We now briefly revisit some of the quantum operations relevant to this work.\\

\noindent {\bf Quantum Entanglement:} A two qubit state is said to be entangled if it can not be written as a product of its individual qubit states. The following four entangled two qubit states are of particular importance.
$$\ket{\phi^{\pm}_{ij}} = \frac{\ket{0_i}\ket{0_j} \pm \ket{1_i}\ket{1_j}}{\sqrt{2}},\;\ket{\psi^{\pm}_{ij}} = \frac{\ket{0_i}\ket{1_j} \pm \ket{1_i}\ket{0_j}}{\sqrt{2}},$$
\noindent We refer to the two qubit entangled states $\ket{\phi^{\pm}_{ij}}$ and $\ket{\psi^{\pm}_{ij}}$ as Bell pairs and $\ket{\phi^{+}_{ij}}$ as an EPR pair shared between node $i$ and node $j.$ Also, we refer to $\ket{\phi^{+}_{0i}}$ and $\ket{\phi^{+}_{ij}}$ as link level entanglement and end-to-end user entanglement respectively.\\

\noindent {\bf Fidelity:} Fidelity captures the closeness between two quantum states. EPR pairs obtained after entanglement may not be perfect due to noise in quantum communication channels and quantum operations. Suppose, nodes $i$ and $j$ share an imperfect EPR pair $\ket{\sigma_{ij}}$. The density operator $\rho_{ij}$ associated with $\ket{\sigma_{ij}}$ can be expressed in the Bell basis as follows.

{\small
\begin{align}
\rho_{ij} = F_1\ket{\phi^{+}_{ij}}\bra{\phi^{+}_{ij}}&+F_2\ket{\psi^{-}_{ij}}\bra{\psi^{-}_{ij}}\nonumber\\
&+ F_3\ket{\psi^{+}_{ij}}\bra{\psi^{+}_{ij}}+F_4\ket{\phi^{-}_{ij}}\bra{\phi^{-}_{ij}},\label{eq:bell-basis}
\end{align}
}\ignorespacesafterend 
with $\sum_{i=1}^4 F_i = 1.$ Since, we are interested in state $\ket{\phi^{+}_{ij}}$, $F_1$ is the fidelity of state $\ket{\sigma_{ij}}$. Ideally one would want to achieve a unit fidelity, i.e. value of $F_1$ to be equal or very close to $1.$ But, due to non-ideal conditions, $F_1 < 1.$\\

\noindent{\bf Entanglement Swapping \cite{Zukowski93}:} Suppose end users $i,j$ share EPR pairs $\ket{\phi^{+}_{0i}},\ket{\phi^{+}_{0j}}$ with the switch. Then an entangled state $\ket{\phi^{+}_{ij}}$ can be created between users $i$ and $j$ by taking a Bell state measurement (BSM) on one qubit of each pair stored at the switch. This operation is known as entanglement swapping. Let $F_{link}$ denote the fidelity of both link level entanglements on which entanglement swapping is being performed. Denote $F_{swap}$ as the fidelity of the new EPR pair created by the swap. Then, $F_{swap}$ can be expressed as an increasing function $\eta(\cdot)$ of $F_{link}$ with $F_{swap} < F_{link}.$ For example, when link level entanglements are assumed to be Werner states, $\eta(F_{link}) = 1/4 + 3/4 ((4F_{link}-1)/3)^2$ \cite{Briegel98}.\\

\noindent{\bf Entanglement Purification \cite{Dur1999}:} The EPR pairs obtained after entanglement swapping may not be perfect and may have low fidelities. Assume that end user $i$ and $j$ share a number $\kappa (\ge 2)$ of imperfect EPR pairs $\ket{\sigma_{ij}}$. However, nodes $i$ and $j$ can perform local gates, measurements on $\kappa$ $\ket{\sigma_{ij}}$ EPR pairs along with classical information exchange to obtain a single higher quality EPR pair. This operation is known as entanglement purification. Note that, entanglement purification is probabilistic. We refer interested readers to \cite{Dur2007} for a more elaborate explanation of the purification procedure. This procedure can be repeated in a nested manner on the newly purified set of EPR pairs until desired fidelity is achieved.\\

\subsection{System Model}
In this Section, we introduce the system model used in the rest of the paper. We summarize notations in Table \ref{table:smstab}. We use the notation $P_X(x)$ to represent the probability that the random variable $X$ takes the value $x$, i.e., $P_X(x) = \Pr[X = x]$. Also, let  $\mathbb{E}[X]$ denote the expected value of random variable $X.$ We use a bold letter symbol, such as $\pmb{V}$, to represent a vector.\\

\noindent{\bf End-to-end User Entanglement Requests:} We assume time is slotted with slot duration $\Delta$ seconds. Let $n$ denote a time slot with $n \in \{0, 1, \cdots\}$. At each time slot, requests to create $\ket{\phi^{+}_{ij}}$ arrive at the switch. Let $A_{ij}(n)$ denote the number of such requests arriving in time slot $n$. We assume $\{A_{ij}(n)\}_{n\ge 0}$ to be mutually independent i.i.d processes with rate $\lambda_{ij}$, i.e. $\mathbb{E}[A_{ij}(n)] = \lambda_{ij}$, $i,j\in\{1,K\}$. We also assume that these requests require an application level minimum target fidelity\footnote{Throughout the paper, the fidelity of a state is always with respect to the ideal EPR pair $\ket{\phi^+}$.} of  $F^{th}$.\\

\noindent{\bf Link Level Entanglement Generation:} At the beginning of each time slot (say $n$), the network infrastructure attempts to create $\alpha_{max}$ number of link level entanglements between an end user $i$ and the switch. We assume the switch and the user has enough quantum memory to store all entanglements. We denote $p_i$ to be the success probability to create a single link level entanglement between user $i$ and the switch in each of these attempts. $p_i = e^{-\theta d_i}$, where $d_i$ is the length of link $i$ and $\theta$ is its attenuation coefficient. Denote $\pmb{p} = [p_{i}, i=1,\cdots,K].$ Clearly, the number of link level entanglements generated between user $i$ and the switch ($T_i(n)$) is a binomial random variable with parameters $\alpha_{max}$ and $p_i$, i.e.,
\[
P_{T_i}(a_{i}) = {\alpha_{max} \choose a_{i}}p_i^{a_{i}}(1-p_i)^{\alpha_{max}-a_i},\quad a_i =0,\ldots ,\alpha_{max}
\]

Let $\mathcal{A}$ denote the set of all feasible vectors of $\pmb{T(n)} = [T_1(n), \cdots, T_K(n)]$, i.e., $\mathcal{A} = \{0,1,2,\cdots, \alpha_{max}\}^K$. We assume the link level entanglements to be valid only for a single time slot and they decohere at the end of the time slot. Denote $F_{link}$ to be the fidelity of a successfully generated link level entanglement. Due to noise induced in the link, $F_{link} < 1$.\\

\noindent{\bf Nested Entanglement Purification:} When the fidelities of the end-to-end entanglements generated after swaps are less than $F^{th}$, the entanglements become unusable for consumption at the application level. To circumvent this issue, one can perform nested entanglement purification on these low quality entanglements to generate entanglements with a target fidelity. Suppose $Y$ denotes the random variable representing the total number of output EPR pairs with fidelity greater than $F^{th}$ produced by a nested purification routine by consuming $X$ low quality EPR pairs. We refer to $P_{Y|X}$ and $\mathbb{E}[Y|X]$ as the conditional yield\footnote{Note that, the definition of yield is slightly different from what has been used in literature. Previous work refers to $Y/X$ as the yield of a purification routine.} distribution and the conditional expected yield of the nested entanglement purification routine. We derive expressions for $P_{Y|X}$ and $\mathbb{E}[Y|X]$ for different classes of nested entanglement purification protocols in Section \ref{expected_yield}. Such a purification can be performed before or after entanglement swapping routine resulting in the following two architectures as shown in Figure \ref{stability} (a) and (b). 
\begin{enumerate}[(a)]
\item {\bf Purify and swap (PS) architecture -} In this architecture, nested entanglement purification is first performed on individual link level entanglements after which entanglement swapping is performed. 
\item {\bf Swap and purify (SP) architecture -} This architecture first performs entanglement swapping on link level entanglements followed by end-to-end purification.
\end{enumerate}
Let $D_{ij}(n)$ ($\tilde{D}_{ij}(n)$) denote the number of high quality end-to-end entanglements that are delivered to the requesting application at time slot $n,$ each with fidelity $F_{final}$ ($\tilde{F}_{final}$) such that $F_{final} \ge F^{th}$ ($\tilde{F}_{final} \ge F^{th}$) in the SP (PS) architecture.\\

\noindent{\bf Entanglement Scheduling:}
At the beginning of each time slot, the switch stores the unserved requests in a queue. Let $Q_{ij}(n)$ ($\tilde{Q}_{ij}(n)$) denote the queue size in time slot $n$ in the SP (PS) architecture. Also, let $\pmb{Q(n)} = [Q_{ij}(n), i=1,\cdots,K \text{ and } j = i+1,\cdots,K]$. Similarly, let $\pmb{\tilde{Q}(n)} = [\tilde{Q}_{ij}(n), i=1,\cdots,K \text{ and } j = i+1,\cdots,K]$. The switch makes scheduling decisions in each time slot and selects $W_{ij}(n)$ ($\tilde{W}_{ij}(n)$) number of link level entanglements to perform entanglement swapping under SP (PS). Since entanglement swapping procedures are probabilistic (with success probability $q_{ij}$), only a fraction of $W_{ij}(n)$ ($\tilde{W}_{ij}(n)$) EPR pairs are created between user $i$ and $j$ at time slot $n$ under SP (PS) architecture. Denote $\pmb{q} = [q_{ij}, i=1,\cdots,K \text{ and } j = i+1,\cdots,K]$, $\pmb{W(n)} = [W_{ij}(n), i=1,\cdots,K \text{ and } j = i+1,\cdots,K]$ and $\pmb{\tilde{W}(n)} = [\tilde{W}_{ij}(n), i=1,\cdots,K \text{ and } j = i+1,\cdots,K]$.

\begin{table}[htbp]
\begin{center}
		\begin{tabular}{ l|l} 
			\hline
			$K$&Number of end users\\
			$F^{th}$&Application level target fidelity threshold\\
			$F_{link}$&Initial link level fidelity\\
			$F_{swap}$ ($\tilde{F}_{swap}$)&Fidelity after a successful ent. swap under SP (PS)\\
			$F_{final}$ ($\tilde{F}_{final}$)&Fidelity of EPR pair delivered to app under SP (PS)\\
			$p_i$&Ent. generation success prob. for link $i$\\
			$q_{ij}$&Ent. swap success prob. between user $i$ and $j$\\
			$\alpha_{max}$&Max. no. of link level ent. generation attempts/slot\\
			$n$& A time slot\\
			$A_{ij}(n)$& No. of requests arrived in slot $n$ for creating ent.\\
			&b/w user $i$ and $j$ with Fidelity $F^{th}$\\
			$\lambda_{ij}$&Avg. request arrival rate to create ent. between \\
			&user $i$ and $j$\\
			$T_{i}(n)$& No. of successful link level ent. created for link $i$\\
			&in slot $n$\\
			$\tilde{T}_{i}(n)$& No. of successful purified link level ent. created\\
			&for link $i$ under PS in slot $n$\\
			$W_{ij}(n)$ ($\tilde{W}_{ij}(n)$)& No. of link level ent. chosen to perform swap\\
			&b/w user $i$ and $j$ under SP (PS) in slot $n$\\
			$Q_{ij}(n)$ ($\tilde{Q}_{ij}(n)$)& No. of pending requests in slot $n$ for creating ent.\\
			&b/w user $i$ and $j$ with Fidelity $F^{th}$ under SP (PS)\\
			$B_{ij}(n)$& No. of successful ent. swaps performed in slot $n$\\
			&b/w user $i$ and $j$ under SP\\
			$D_{ij}(n)$ ($\tilde{D}_{ij}(n)$)& No. of successful ent. created in slot $n$ b/w \\
			&user $i$ and $j$ with fidelity $> F^{th}$ under SP (PS)\\
			$\mathbb{E}[Y|X]$&Conditional expected yield of an ent. pur. routine\\
			$P_{Y|X}$&Conditional yield distribution of an ent. pur. routine\\
			$\eta(\cdot)$& Fidelity after ent. swap as a function of link fidelity\\
			\hline
		\end{tabular}
\end{center}		
		\caption{Summary of Notations.}
		\label{table:smstab}
\end{table}
\section{Capacity Region Computation: Main Results}\label{capacity}
In this section, we present the main results related to the capacity region computation of a switch under both PS and SP architecture. We also propose corresponding max weight scheduling policies that stabilize the switch for all feasible request arrival rates. We relegate the proofs in this section to the Appendix \ref{sec:appendix}.

\begin{definition}
A switch is defined to be stable under a scheduling policy if the process $\{\pmb{Q(n)}\}_{n=0}^\infty$ converges in distribution to $\pmb{Q}$ independent of the initial condition and $\mathbb{E}[\pmb{Q}] < \infty$ \cite{Tassiulas97}.
\end{definition}
\begin{definition}
The capacity region $\Lambda$ of a switch is defined to be the set of request rates $\pmb{\lambda}$ for which the switch remains stable. 
\end{definition}
\subsection{PS Architecture}
In this architecture, entanglement purification is first performed on link level entanglements, followed by entanglement swap. We denote $\tilde{T}_{i}(n)$ and $\tilde{F}_{link}$ to be the number of link-level entanglements and their fidelities  after link-level purification is performed. Since, the purification is on the link level, we need to identify a target threshold for fidelity on link-level ($F_{link}^{th}$). We set $F_{link}^{th} = \eta^{-1}(F^{th})$. We define the probability distribution of $\tilde{T}_{i}(n)$ as 
\begin{align}
P_{\tilde{T}_{i}}(\tilde{a}_{i}) = \sum\limits_{a_i \ge \tilde{a}_{i}}^{\alpha_{max}}P_{Y|X}(\tilde{a}_{i}|a_{i})P_{T_{i}}(a_{i})\label{eq:tilde}
\end{align}
\noindent where $P_{Y|X}$ is the conditional yield distribution of the purification protocol. The following theorem computes the capacity region under PS architecture. 
\begin{theorem}\label{ps:capacity}
The capacity region of the switch under PS is given by,
\begin{align}\label{eq:capacity}
    &\Lambda^{PS}(\alpha_{max}, \pmb{p}, F, F^{th}, \pmb{q}, P_{Y|X}) \nonumber \\
    &= \bigg\{\pmb{\lambda}:\exists b_{\pmb{\tilde{a}},\pmb{\pi}} \text{ s.t. } \pmb{\lambda} \preceq \sum\limits_{\pmb{\tilde{a}}\in \mathcal{A}, \pmb{\tilde{a} \neq 0}}P_{\pmb{\tilde{T}}}(\pmb{\tilde{a}})\sum\limits_{\pmb{\pi}:\sum\limits_{i} \pi_{ij} \le \tilde{a_j}} b_{\pmb{\tilde{a}},\pmb{\pi}}\pmb{q}\pmb{\pi},\nonumber\\
    & b_{\pmb{\tilde{a}},\pmb{\pi}} > 0, \sum\limits_{\pi}b_{\pmb{\tilde{a}},\pmb{\pi}} < 1\; \forall \pmb{\tilde{a}} \in \mathcal{A}\bigg\},
\end{align}
\noindent where $c \preceq d$ is defined as the component-wise inequality between vectors $c$ and $d.$
\end{theorem}

Intuitively, $b_{\pmb{\tilde{a}},\pmb{\pi}}$ can be interpreted as the fraction of time in which a scheduling policy selects $\pmb{\pi}$ given $\pmb{\tilde{a}}$. Similarly, $\pmb{q}\pmb{\pi}$ is the element-wise multiplication of the vectors $\pmb{q}$ and $\pmb{\pi}$. $q_{ij}\pi_{ij}$ can be thought of as the average number of successfully served entanglement requests between users $i$ and $j$ in time slot $n$. Note that, the Noise Less capacity region, i.e., the case when no purification is applied and there is no notion of $F^{th},$ can be obtained by setting $P_{\pmb{\tilde{T}}}(\pmb{\tilde{a}}) = P_{\pmb{T}}(\pmb{\tilde{a}})$. This is precisely what authors in \cite{Vasantam22} have obtained for the case when $\alpha_{max}=1.$ We now define the following max-weight scheduling policy for the PS architecture.

\begin{definition} {\bf (MW-PS)} Under this policy, the switch selects the purified link level entanglements from the available entanglements ($\tilde{a}_j(n)$) in the following manner to perform entanglement swaps at each time slot $n$.
\begin{align}
\pmb{\tilde{W}(n)} = &\argmax_{\pi} \sum\limits_{i,j} \pi_{ij}q_{ij}\tilde{Q}_{ij}(n)\nonumber\\
&\text{s.t}\quad \sum\limits_{i} \pi_{ij} \le \tilde{a}_j(n),\quad \forall j
\end{align}
\end{definition}
The theorem below discusses the stability properties of MW-PS.
\begin{theorem}\label{ps:mw}
If $\mathbb{E}[A_{ij}^2(n)]$ is finite $\forall i, j$, then the capacity region of MW-PS coincides with $\Lambda^{PS}.$
\end{theorem}
\subsection{SP Architecture}
In this architecture, entanglement swapping is performed first, followed by end-to-end purification. We characterize its capacity region as follows.
\begin{theorem}\label{sp:capacity}
The capacity region of a quantum switch under SP is given by 
\begin{align}\label{eq:capacity}
    &\Lambda^{SP}(\alpha_{max}, \pmb{p}, F, F^{th}, \pmb{q}, \mathbb{E}[Y|X]) \nonumber\\
    &= \bigg\{\pmb{\lambda}:\exists b_{\pmb{a},\pmb{\pi}} \text{ s.t. } \pmb{\lambda} \preceq \sum\limits_{\pmb{a}\in \mathcal{A}, \pmb{a}\neq 0}P_{\pmb{T}}(\pmb{a}) \sum\limits_{\pmb{\pi}:\sum\limits_{i} \pi_{ij}(n) \le a_j} b_{\pmb{a},\pmb{\pi}} \hat{H}(\pmb{\pi}),\nonumber\\
    & b_{\pmb{a},\pmb{\pi}} > 0, \sum\limits_{\pi}b_{\pmb{a},\pmb{\pi}} < 1\; \forall \pmb{a}\in \mathcal{A}\bigg\},
\end{align}
\noindent Also, $\hat{H}(\cdot)$ is given by the following expression.
$$\hat{H}(\pi_{ij}) = \sum\limits_{l=0}^{\pi_{ij}} \mathbb{E}[Y|X=l] {\pi_{ij} \choose l}q_{ij}^l(1-q_{ij})^{\pi_{ij}-l},$$
\noindent Here, $\mathbb{E}[Y|X=l]$ is the conditional expected yield value of a purification protocol.
\end{theorem}

The quantity $\hat{H}(\pi_{ij})$ can be interpreted as the average number of successfully served requests in time slot $n$ after purification. We now define a max weight scheduling policy corresponding to SP.
\begin{definition} {\bf (MW-SP)}  In this policy, the switch selects the link level entanglements from the set of available link entanglements ($a_j(n)$) in the following manner to perform entanglement swap at each time slot $n$.
\begin{align}
W(n) = &\argmax_{\pi} \sum\limits_{i,j} \hat{H}(\pi_{ij})Q_{ij}(n)\nonumber\\
&\text{s.t}\quad \sum\limits_{i} \pi_{ij} \le a_j(n),\quad \forall j
\end{align}
\noindent 
\end{definition}
\begin{theorem}\label{sp:mw}
If $\mathbb{E}[A_{ij}^2(n)]$ is finite $\forall i, j$, then the capacity region of MW-SP coincides with $\Lambda^{SP}.$
\end{theorem}
\subsection{Boundary of the Capacity Region and Maximum Throughput}
We now formulate a weighted throughput maximization problem to determine the boundary of $\Lambda^{PS}$ and $\Lambda^{SP}$ as follows. Given a quantum switch, we want to find the maximum weighted achievable user request rate that stabilizes it. Let $w_{ij} \ge 0$ denote the weight associated with end user-pair $(i, j)$. Formally, we define the following optimization problem for PS:
\begin{align}
    &\max \quad \lambda\nonumber \\
    \text{s.t} \quad &[w_{12}\lambda, \cdots, w_{K-1K}\lambda] \in \Lambda^{PS}(\alpha_{max}, \pmb{p}, F, F^{th}, \pmb{q}, P_{Y|X}).\label{eq:cap_max}
\end{align}
A similar optimization problem can be formulated for SP architecture as well. One application of the above optimization problem is to sketch out the capacity region of the switch by solving it for different instances of $[w_{12}, \cdots, w_{K-1K}]$, as each solution falls on the boundary of the capacity region. Note that, both the objective and constraints in \eqref{eq:cap_max} are linear and involves continuous decision variables. Hence optimization problem \eqref{eq:cap_max} corresponds to a linear program.

\begin{figure*}[htbp]
\centering
\begin{minipage}{0.5\textwidth}
\includegraphics[width=1\textwidth]{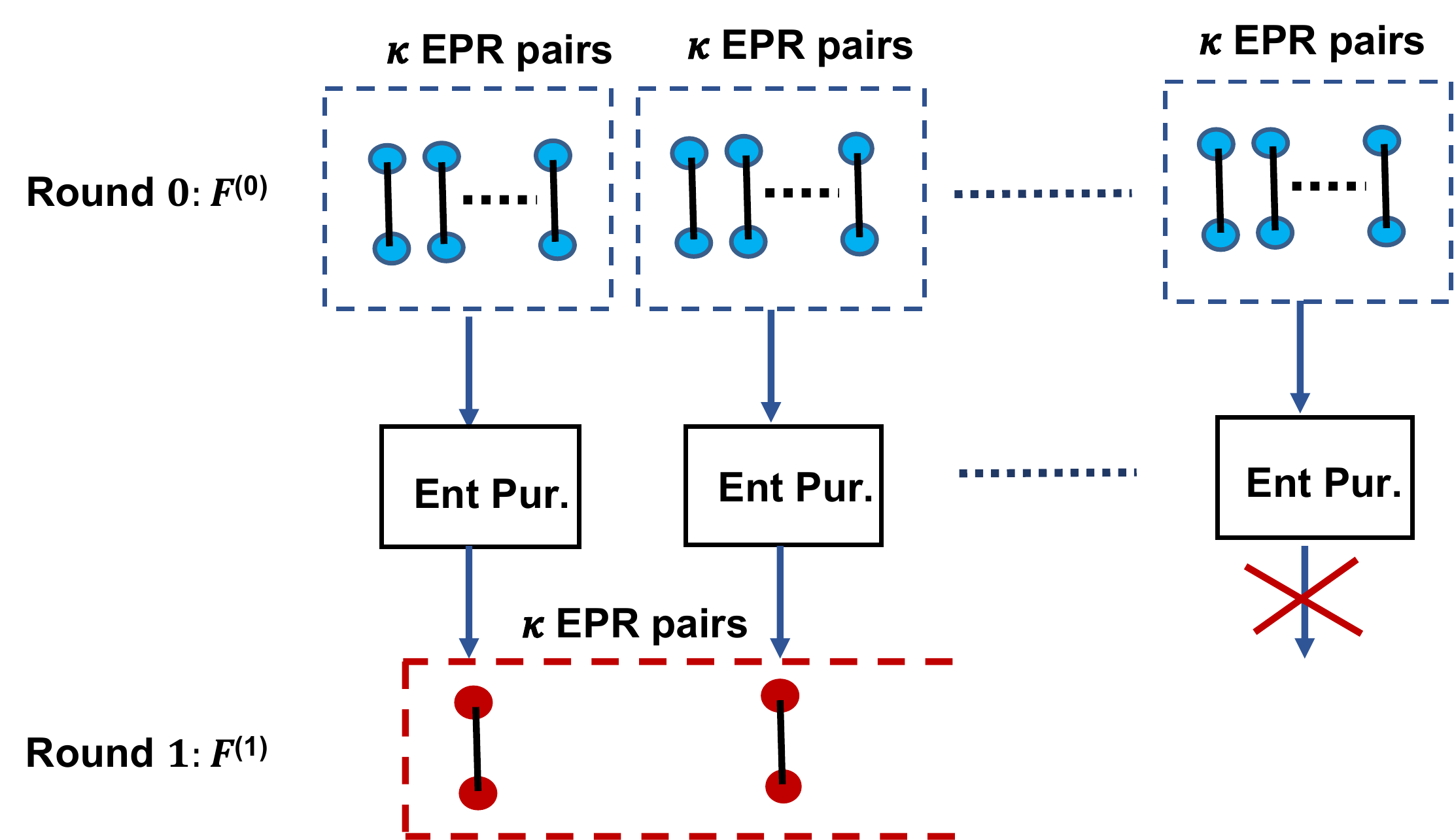}
\subcaption{}
\end{minipage}
\hspace{1cm}
\begin{minipage}{0.4\textwidth}
\includegraphics[width=1\textwidth]{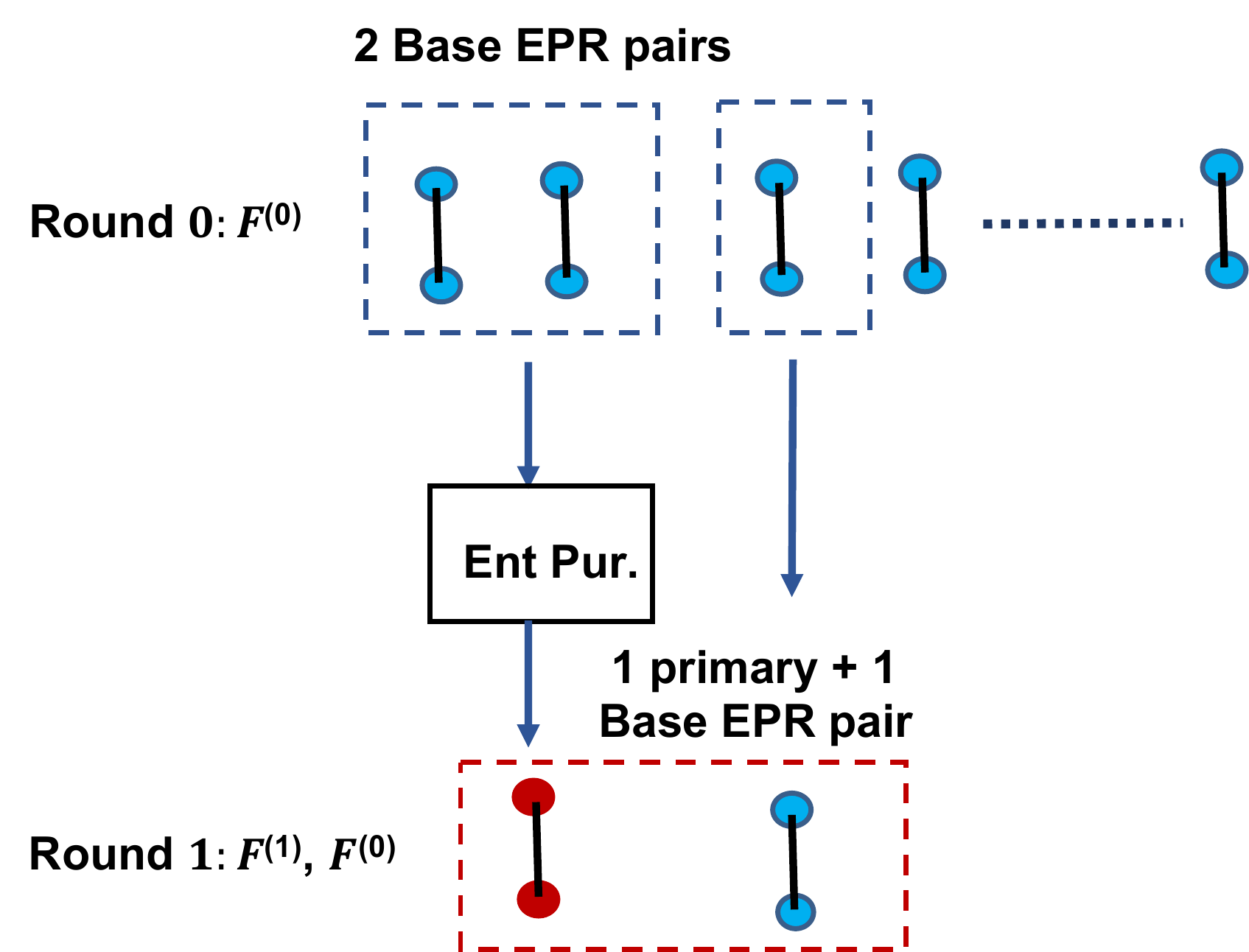}
\subcaption{}
\end{minipage}\hfill
\caption{(a) Symmetric and (b) Entanglement Pumping Purification Protocols.}
\label{purification-types}
\end{figure*}
\section{Properties of Specific Nested Entanglement Purification Protocols} \label{expected_yield}

\begin{table*}[h]
	\begin{center}
		\begin{tabular}{ | c| c| c| c| c|} 
			\hline
			Protocol& Input State& $\kappa$& $\omega_S(F)$& $\chi_S(F)$\\ 
			\hline
			DEJMPS \cite{Deutsch96}& Bell-Diagonal (See \eqref{eq:bell-basis} with $F_1=F$)&$2$&$(F+F_2)^2 + (F_3+F_4)^2$&$\frac{F^2+F_2^2}{(F+F_2)^2 + (F+F_4)^2}$\\
			DEJMPS \cite{Deutsch96}& Binary&$2$&$F^2 + (1-F)^2$&$\frac{F^2}{F^2 + (1-F)^2}$\\
			BBPSSW \cite{Briegel98}& Werner&$2$&$\bigg(F + \frac{1-F}{3}\bigg)^2 + \bigg(\frac{2(1-F)}{3}\bigg)^2$&$\frac{F^2 + \frac{1}{9}(1-F)^2}{F^2 + \frac{2}{3}F(1-F) + \frac{5}{9}(1-F)^2}$\\
			\hline
		\end{tabular}
		\caption{Success probability and output fidelity for different symmetric Purification protocols.}
		\label{table:spp}
	\end{center}
\end{table*}

We now derive the conditional yield distribution and conditional expected yield of the following classes of nested entanglement purification protocols. Let $X$ denote the random variable representing the total number of input EPR pairs submitted for purification to a nested purification routine. Similarly, let $Y$ denote the random variable representing the the total number of output EPR pairs with fidelity greater than $F^{th}$ produced by the nested purification routine. We are interested in $P_{Y|X}$ and $\mathbb{E}[Y|X]$.

\subsection{Symmetric Purification Protocols}\label{sub:symm}
In symmetric purification protocols, purification is performed on identical EPR pairs, i.e., all input EPR pairs have equal fidelities. The protocol is often performed in multiple rounds, hence also known as recurrence protocol \cite{Dur1999}. At each purification round, the total number of available EPR pairs are divided into groups of $\kappa > 1$ number of EPR pairs as shown in Figure \ref{purification-types} (a). Purification is then performed on each group. If the purification is successful, then only one EPR pair out of $\kappa$ EPR pairs are kept in each group. If the purification is unsuccessful, all of the $\kappa$ EPR pairs are discarded. The remaining EPR pairs are used for purification at the next round.

At each round, fidelities of EPR pairs increase by a certain fraction after purification. Purification is continued until either there are no EPR pairs left or the application level fidelity threshold is reached for the output EPR pairs. We denote $L$ as the maximum number of purification rounds.

Let $Y^{(l)}$ and $F^{(l)}$ denote the total number of output EPR pairs and their fidelities after $l$ rounds of purification for $l = 1,2, \cdots, L$. Thus we have  $F^{(L)} \ge F^{th}$ but $F^{(L-1)} < F^{th}.$ We also have $Y^{(L)} = Y$ and $Y^{(0)} = X.$

Let $x$ denote the total number of input EPR pairs, each with fidelity $F$, available at the beginning for purification. At round $l$, each purification unit takes $\kappa$ number of EPR pairs each with fidelity $F^{(l-1)}$ and produce one EPR pair of fidelity $F^{(l)} = \chi_S(F^{(l-1)})$ with $F^{(l-1)} < F^{(l)}$ and $l\in\{1, 2, \cdots, L\}$, $F^{(0)} = F$. Here, $\chi_S(\cdot)$ is the output fidelity function.

Purification is probabilistic and its success probability depends on the fidelity of the input EPR pairs. Let $r_S^{(l)}$ denote the success probability of a purification routine in round $l$. Thus we have

 \begin{align}
    r_S^{(l)} = \omega_S(F^{(l-1)})\quad \forall l\in\{1, 2, \cdots, L\},
\end{align}

\noindent where $\omega_S(\cdot)$ denotes the success probability as a function of fidelity of input EPR pairs.

The probability distribution of number of output EPR pairs produced after $l$ rounds of purification can be derived by the following recursion with $\forall l\in\{2, \cdots, L\}$.
\begin{align}
    &P_{Y^{(l)}|X}(y|x)\\
    &= \sum\limits_{z=y*\kappa}^{\lfloor x/\kappa^{l-1}\rfloor}{\lfloor z/\kappa\rfloor \choose y} \bigg(r_S^{(l)}\bigg)^{y}\bigg(1-r_S^{(l)}\bigg)^{\lfloor z/\kappa\rfloor-y} \nonumber\\
    &\quad\quad\quad\quad\quad\quad\quad\quad\quad\quad\quad \quad\quad \cdot P_{Y^{(l-1)}|X}(z|x),\label{eq:prob_dist_symm}
\end{align}

\noindent with $P_{Y^{(1)}|X}(y|x) = {\lfloor x/\kappa\rfloor \choose y}(r_S^{(1)})^{y}(1-r_S^{(1)})^{\lfloor x/\kappa\rfloor-y}$. The conditional yield distribution is given by $P_{Y|X}(y|x) = P_{Y^{(L)}|X}(y|x)$. The conditional expected yield function is given by $    \mathbb{E}[Y|X = x] = \sum_{y=1}^{\lfloor x/\kappa^L\rfloor}y* P_{Y^{(L)}|X}(y|x).$

We present the success probability and output fidelity functions of different protocols that perform $\kappa:1$ purification in Table \ref{table:spp}. Note that, depending on the input state of an EPR pair, $\omega_S(\cdot)$ and $\chi_S(\cdot)$ may be different. Consider the density operator representation of an input EPR pair as shown in Equation \eqref{eq:bell-basis}. When $F_1 = F$ and $F_2 = F_3 = F_4 = (1-F)/3,$ we call the entangled state a Werner state (White noise). When $F_1 = F$ and only one of $F_2, F_3$ or $F_3$ is non-zero, i.e., equals to $(1-F)$, the state is called a Binary state. For example: if the channel has bit flip errors, then a particular kind of binary state ($F_1 = F, F_2= F_4 =0, F_3 = 1-F$) is generated. 
\begin{remark}
The expressions for success probability and output fidelity functions presented in Table \ref{table:spp} assume no quantum gate errors or measurement errors. However, these expressions can be easily modified to account for such imperfections \cite{Briegel98 ,Fujii09}.
\end{remark}
\subsection{Entanglement Pumping}
Entanglement Pumping \cite{Dur1999} is an asymmetric purification protocol where one EPR pair is purified using multiple auxiliary EPR pairs until a desired target fidelity is achieved. Then the process repeats itself to produce more high quality EPR pairs until all auxiliary EPR pairs are exhausted. The protocol is shown in Figure \ref{purification-types} (b). We call the EPR pair that goes under purification as {\it primary} and  other auxiliary pairs as {\it base} EPR pairs.

We start with a given number of base EPR pairs with Fidelity $F.$ We select the first base EPR pair to be a primary EPR pair and purify it using the second base EPR pair. If the purification is successful, the fidelity of the primary EPR pair increases by a fraction and the base EPR pair gets sacrificed. The primary EPR pair then repeatedly gets purified by sacrificing other base EPR pairs until its fidelity is above $F^{th}$. If a purification fails at any step, the next primary EPR pair is chosen from the set of remaining base EPR pairs and the purification process repeats from the beginning. 

Similar to the definitions in Section \ref{sub:symm}, let $F^{(l)}$ denote the fidelity of a primary EPR pair after $l$ rounds of purification. Suppose $L$ rounds of purification is needed for a primary EPR pair to achieve the target fidelity. Since entanglement pumping is asymmetric, the fidelities of input EPR pairs for a single purification operation are different. We denote $\chi_A(F^{(k)}, F)$ and $\omega_A(F^{(k)}, F)$ to be the output fidelity of a primary EPR pair and purification success probability functions respectively. Thus we have, $F^{(l)} = \chi_A(F^{(l-1)}, F).$ We denote $r_A^{(l)}$ to be the purification success probability for round $l,$ i.e. $r_A^{(l)} = \omega_A(F^{(l-1)}, F).$ 

We use the theory of discrete-time renewal processes \cite{Feller} to derive the distribution $P_{Y|X}$ for entanglement pumping protocol as follows. Henceforth, we will call the purified primary EPR pair with fidelity greater than $F^{th}$ as a high quality EPR pair. We refer to the event of generating a high quality EPR pair as a renewal event. We sequentially select a base pair to purify the primary EPR pair and the ids of base EPR pairs can be mapped to (discrete) time in a renewal process. The number of EPR pairs sacrificed to generate one high quality EPR pair can be considered as the renewal inter-occurrence time. Let $\tilde{X}_i, i = 1, 2, \cdots,$ denote such renewal inter-occurrence times with distribution $P_{\tilde{X}}$. Thus, by the theory of renewal processes \cite{Feller}, the conditional expected yield is given by the following recursion.
\begin{align}
    \mathbb{E}[Y|X = x] = \sum\limits_{i=1}^{x} [1 + \mathbb{E}[Y|X = x-i]]P_{\tilde{X}}(i),
\end{align}
\noindent with $\mathbb{E}[Y|X = j] = 0, \forall j \in \{0,1, \cdots, L\}$. 

Now, let $\tilde{S}_k, k = 1, 2, \cdots,$ denote renewal event occurrence times, i.e. $ \tilde{S}_k= \sum_{i=1}^{k} \tilde{X}_i.$ $P_{\tilde{S}_k}$ can be found out by $k-$fold convolution using the recursion, $P_{\tilde{S}_k}(s) = \sum_{i=1}^{s}P_{\tilde{X}}(i)P_{\tilde{S}_{k-1}}(s-i).$ The conditional yield distribution $P_{Y|X}$ can be computed as
\begin{align}
P_{Y|X}(y|x) = \sum\limits_{i=1}^{x}P_{\tilde{S}_y}(i) - \sum\limits_{i=1}^{x}P_{\tilde{S}_{y+1}}(i).
\end{align}
Finally, the distribution $P_{\tilde{X}}$ can be computed by the following recursion.
\begin{align}
P_{\tilde{X}}(i) = \prod\limits_{l=1}^{L}r_A^{(l)}\bigg[1 -\sum\limits_{k=1}^{i-L-1} P_{\tilde{X}}(k)\bigg]\cdot\mathbbm{1}_{i\notin I},
\end{align}
\noindent Here, $P_{\tilde{X}}(i) = 0, \forall i \in \{0,1, \cdots, L\}$. The set $I$ in the indicator function denotes the infeasibility set of a purification scheme. For example, for a $2:1$ purification scheme with $L=1$, $I = \{i|i\%2 \neq 0\},$ i.e. the set of odd numbers.
\section{Numerical Results}\label{performance}
\begin{figure*}[htbp]
\centering
\begin{minipage}{0.33\textwidth}
\includegraphics[width=1\textwidth]{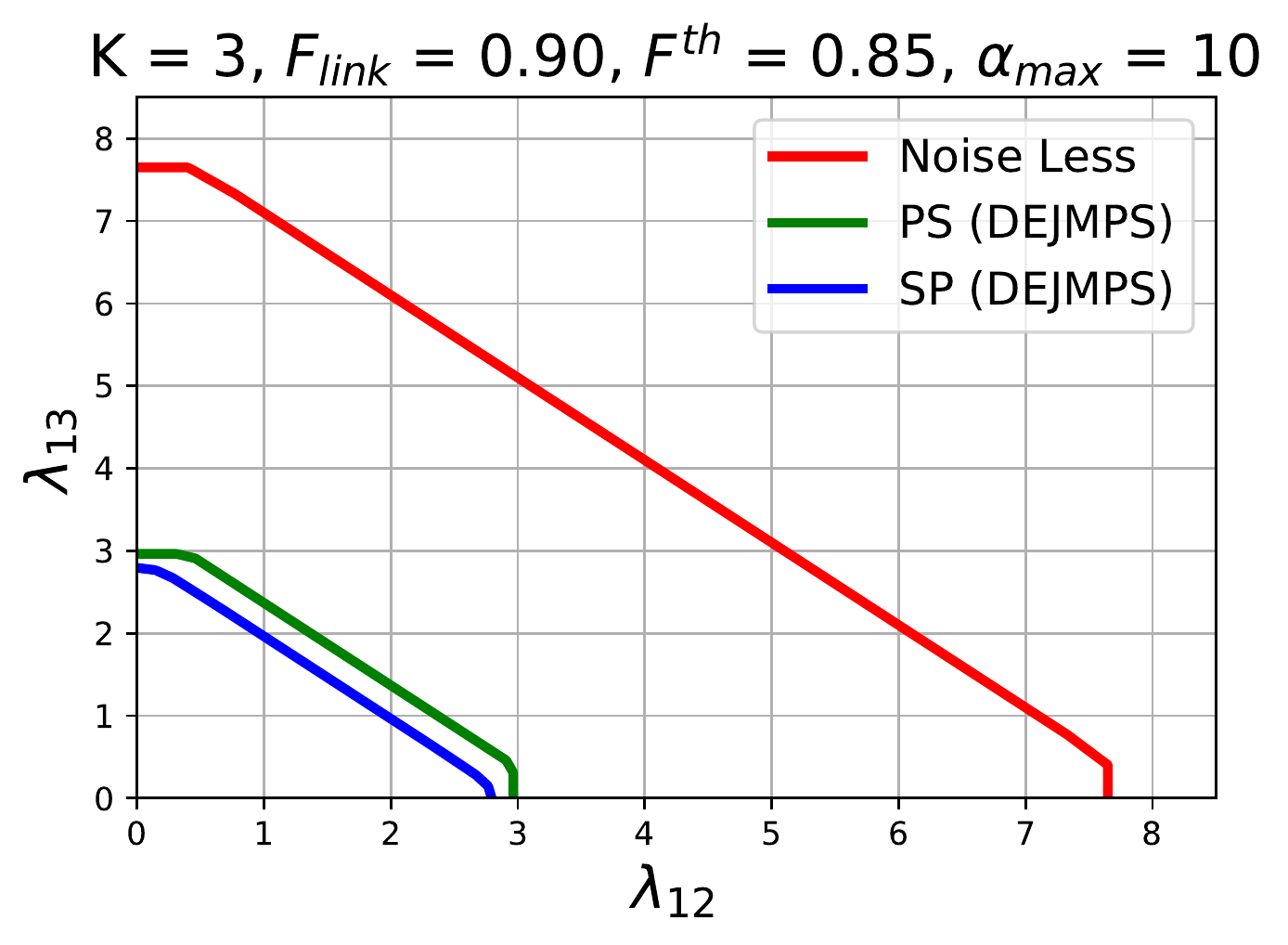}
\subcaption{}
\end{minipage}
\begin{minipage}{0.33\textwidth}
\includegraphics[width=1\textwidth]{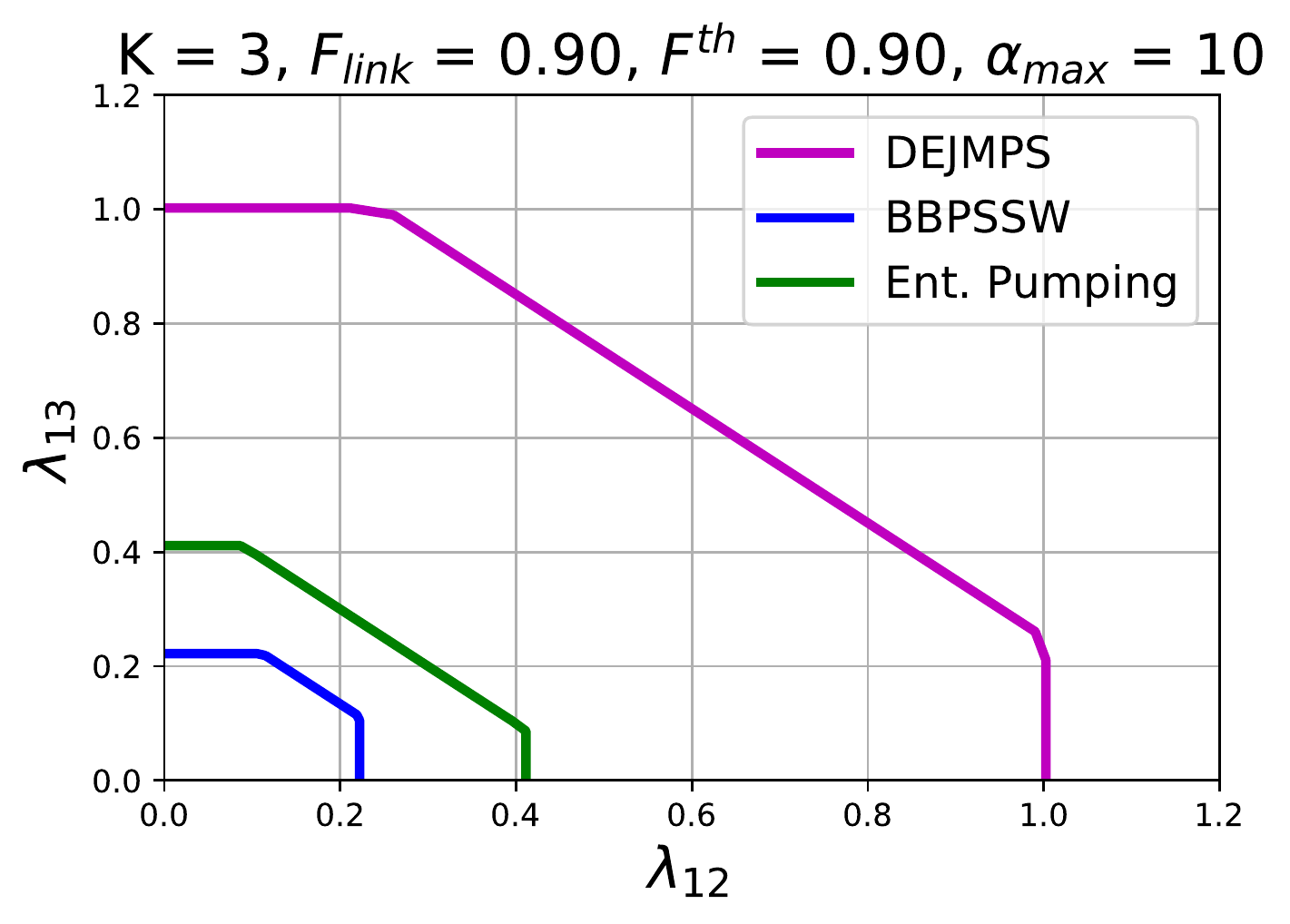}
\subcaption{}
\end{minipage}\hfill
\begin{minipage}{0.33\textwidth}
\includegraphics[width=1\textwidth]{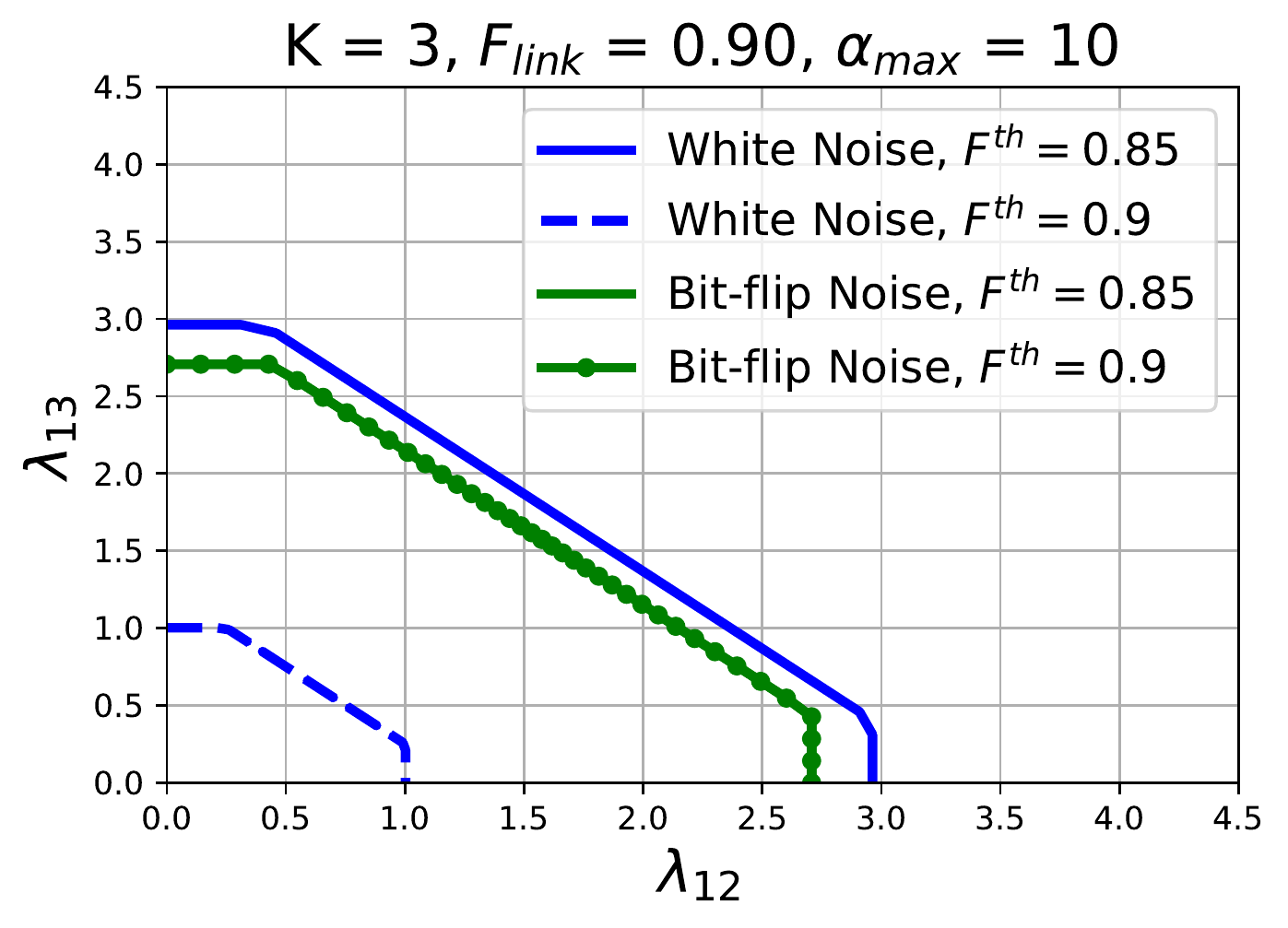}
\subcaption{}
\end{minipage}\hfill
\begin{minipage}{0.33\textwidth}
\includegraphics[width=1\textwidth]{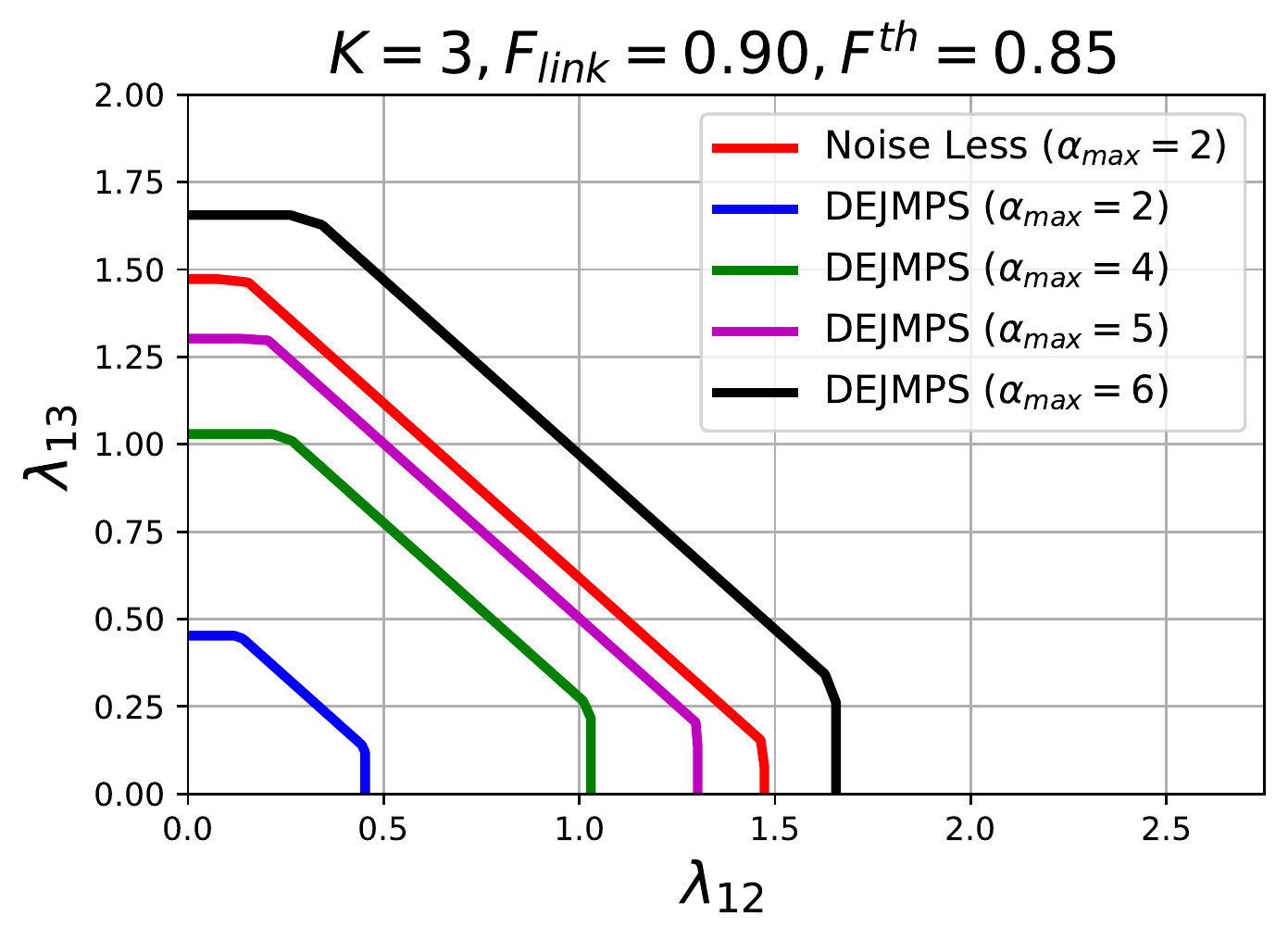}
\subcaption{}
\end{minipage}
\begin{minipage}{0.33\textwidth}
\includegraphics[width=1\textwidth]{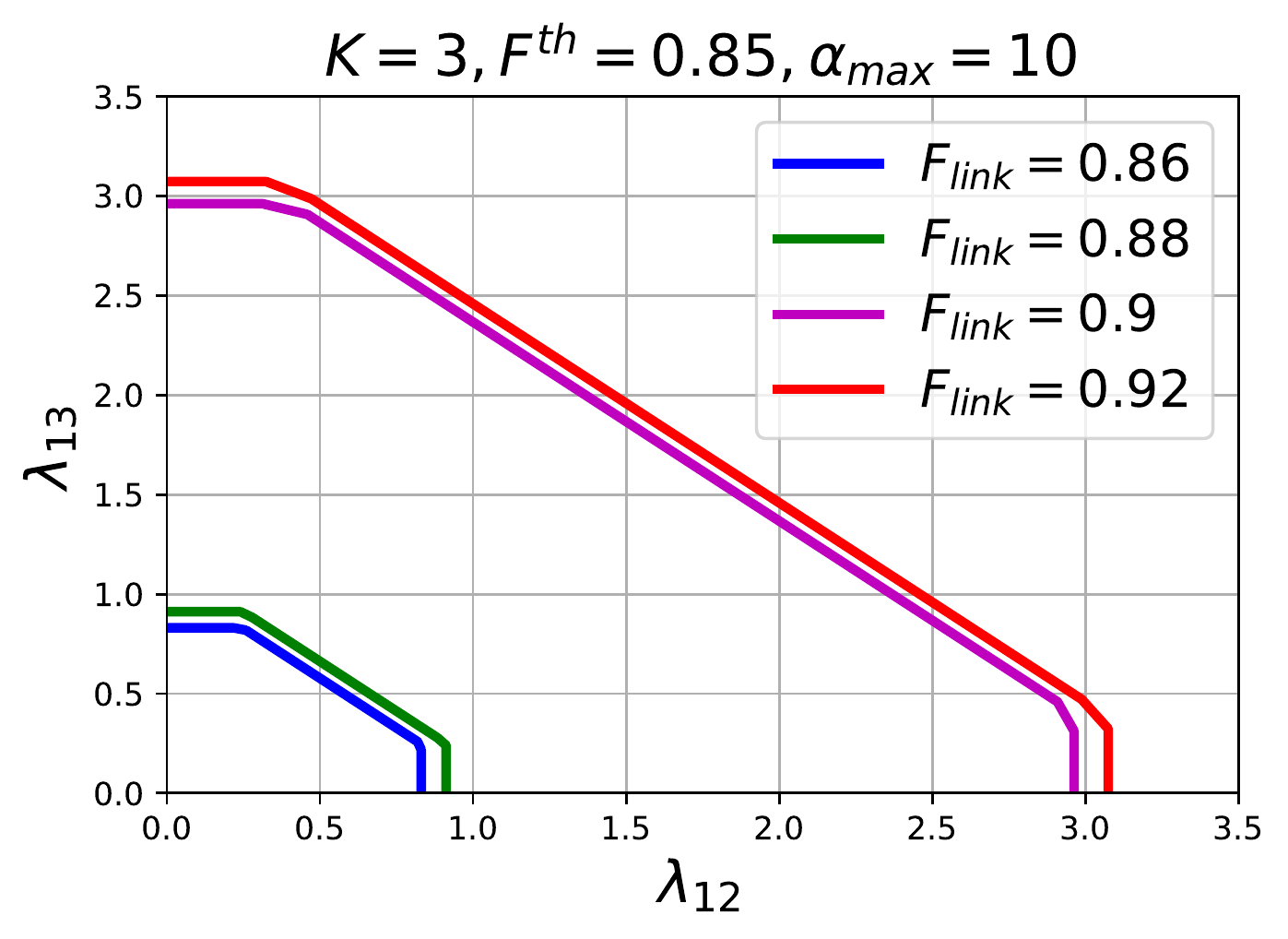}
\subcaption{}
\end{minipage}\hfill
\begin{minipage}{0.33\textwidth}
\includegraphics[width=1\textwidth]{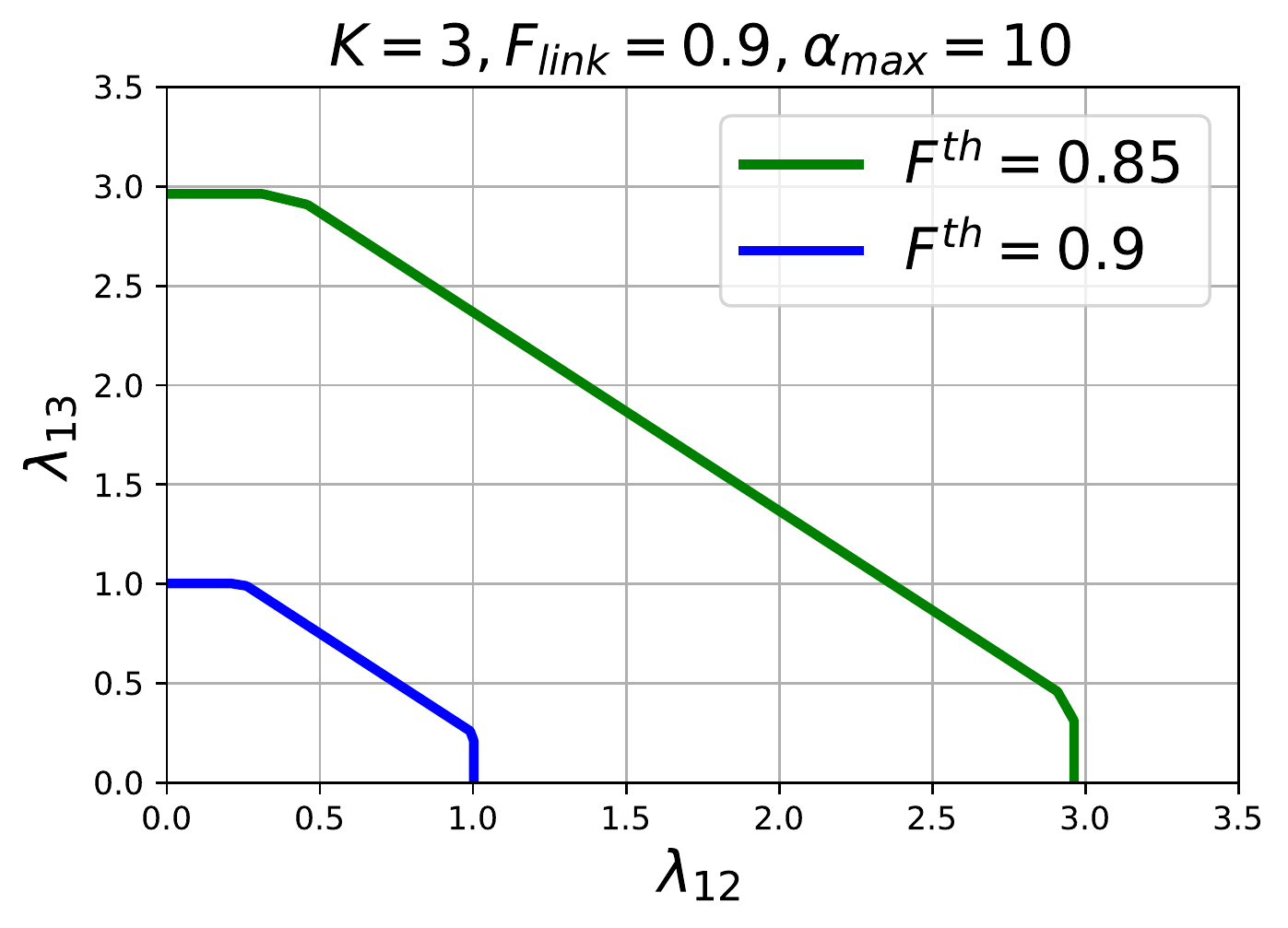}
\subcaption{}
\end{minipage}\hfill
\caption{Capacity Regions computed for different architectures, network parameters and purification protocols. (c)-(f) are computed for PS architecture and DEJMPS purification protocol.}
\label{sim-results}
\end{figure*}
In this section we perform a numerical study to compare the performance of different architectures and purification protocols. In order to obtain the boundary of the capacity region, we solve optimization problem \eqref{eq:cap_max} using the IBM CPLEX solver. The goals of this study are to (i) compare the performance of PS and SP architectures, (ii) to determine which purification protocol works best with respect to switch stability, and (iii) the effect of different network and application-level parameters on the capacity region of the switch. We first present the experimental setup and then discuss the results.
\begin{table}[htbp]
\begin{center}
		\begin{tabular}{ l|lllll l} 
			\hline
			Parameter&$K$&$\alpha_{max}$&$p_i$&$q_{ij}$&$F_{link}$&$F^{th}$\\
			\hline
			Value&$3$&$10$&$0.9$&$0.9$&$0.90$&$0.85$\\
			\hline
		\end{tabular}
\end{center}		
		\caption{Parameters used for numerical studies.}
		\label{table:sim}
\end{table}

We consider three end users connected to a quantum entanglement distribution switch with request arrival vector $\lambda = [\lambda_{12}, \lambda_{13}]$ and $\lambda_{23} = 0.$ We assume the distance between the switch and an end user to be $2.3$ km with fiber attenuation coefficient 0.2 dB/km \cite{Dai21}. Unless specified, we use the parameters presented in Table \ref{table:sim} for numerical simulations. 

\noindent{\bf Comparison of PS and SP architectures:} We compare the capacity region of PS and SP under the DEJMPS protocol as shown in Figure \ref{sim-results} (a). We observe that PS has a larger capacity region compared to SP. This advocates that the switch under PS is stable for a wider range of end-user entanglement generation demands and thus, more robust to unexpected demand fluctuations. We obtain similar results for other purification protocols listed in Table \ref{table:spp} and hence, we omit them. Hence forth, for all remaining experiments, we will use the PS architecture to determine the capacity regions. We also plot the Noise Less capacity region, which serves as an upper bound on the maximum capacity that can be achieved by any purification scheme under any architecture. The Noise Less capacity region is obtained for the setting where there is no notion of $F^{th}$ and links as well as quantum operations are assumed to be perfect. Note that, the Noise Less capacity is significantly higher than those achieved by any of the architectures. The gap between Noise Less capacity and capacities of purification architectures in Figure \ref{sim-results}(a) suggests that there could be room for improvement in purification protocol performance.

\noindent{\bf Comparison of Purification Protocols:} We now compare the capacity regions achieved by three purification protocols: two symmetric (DEJMPS, BBPSSW) and one asymmetric (Pumping). In entanglement pumping, the functions $\chi_A(F^{(k)}, F)$ and $\omega_A(F^{(k)}, F)$ are computed using an asymmetric version of DEJMPS protocol \cite{Dur1999}. We set $F^{th} = 0.9$. In Figure \ref{sim-results} (b), we observe that DEJMPS protocol outperforms the other protocols. DEJMPS requires fewer purification rounds ($L$) than BBPSSW to reach a target fidelity. As the number of base EPR pairs that are sacrificed grows exponentially with $L$, DEJMPS has a larger capacity region. Note that, in \cite{Filip18}, DEJMPS was found to be the optimal two-qubit distillation protocol for Bell-diagonal states. Our results presented in Figure \ref{sim-results} (b) corroborates the optimality result.

\noindent{\bf Bit-flip Noise versus White Noise:} We compare the performance of DEJMPS protocol under two different noise models:  Bit-flip noise and White noise as shown in Figure \ref{sim-results} (c). Short descriptions of both models is found in Section \ref{sub:symm}. We consider two settings: $F^{th} = 0.85$ and $F^{th} = 0.9$. When $F^{th} = 0.85$, both noise models require one ($L=1$) purification round. In this setting, DEJMPS under White noise produces lower quality EPR pairs with higher success probability compared to Bit-flip model. Since, we are only interested in the yield, the success probability component dominates in capacity region calculations. Thus DEJMPS under White Noise provides a slightly larger capacity region as compared to that of under Bit-flip noise. However, for a more stringent requirement on $F^{th}$ ($F^{th} = 0.9$), $L = 1$ and $2$ for Bit-flip and White noise models respectively. The number of EPR pairs that are sacrificed grows exponentially with $L$, resulting in a significantly smaller capacity region under White noise model. We conclude that White noise is difficult to deal with under stringent application level quality requirements. 

\noindent{\bf Effect of $\alpha_{max}$:} In Figure \ref{sim-results}(d), we study the effect of $\alpha_{max}$ (number of link-level entanglement attempts) on the capacity region of the switch. We compute the Noise Less capacity for $\alpha_{max} = 2$ and compare it to the noisy capacity for different values of $\alpha_{max},$ starting with $\alpha_{max} = 2.$ We observe that the Noise Less capacity is significantly larger than the noisy capacity for the same value of $\alpha_{max}$. As expected, increasing $\alpha_{max}$ increases the noisy capacity. We also observe that one has to make at least three times more link-level entanglement attempts (black curve) to achieve the noise-less capacity.

\noindent{\bf Effect of $F_{link}$ and $F^{th}$:} We study the effect of $F_{link}$ and $F^{th}$ on the capacity region of the switch in Figures \ref{sim-results} (e) and (f) respectively. With an increase in the quality of initial link-level entanglements, larger capacity regions are obtained as shown in Figure \ref{sim-results} (e). Note that, the improvement in performance is minute between $F_{link} = 0.86$ and $0.88$ but significant between $F_{link} = 0.88$ and $0.9$. This can be explained by the fact that $L=2$ when $F_{link} = 0.86, 0.88$ and $L=1$ when $F_{link} = 0.9, 0.92.$ Thus the transition from $F_{link} = 0.88$ to $F_{link} = 0.9$ involves one less number of purification round resulting in significant gain in terms of capacity. We also compare the capacity regions of the switch for different values of $F^{th}$ in Figure \ref{sim-results} (f). We observe that the capacity region reduces drastically with stringent application level quality requirements.
\section{Conclusion and Open Problems}\label{conclusion}
In this work, we analyzed the effect of channel noise and entanglement purification on the capacity region of a quantum switch. Going further, we would like to extend our analysis to an arbitrary network setting. Conceptually, this seems doable provided that there is a predefined path in place for each user pair. However, the number of optimization variables in \eqref{eq:cap_max} can grow rapidly as a function of numbers of nodes and user pairs. Also, the focus of this work has been on a swap-purify architecture and a purify-swap architecture. It would be interesting to consider an alternate purify-swap-purify architecture where link-level entanglements are purified followed by swaps followed by end-to-end purification. For the network setting, one can of course perform multi-hop purification, then do swapping followed by again multi-hop purification.
\section{Acknowledgments}
This research was supported by the NSF Engineering Research Center for Quantum Networks (CQN), awarded under cooperative agreement number 1941583 and by NSF grant CNS-1955834.
\section{Appendix}\label{sec:appendix}
The proofs of Theorems presented in Section \ref{capacity} are in spirit similar to the proofs presented in \cite{Vasantam22, Tassiulas97}. In particular, in \cite{Vasantam22}, authors derived the Noise Less capacity region of a quantum switch with $\alpha_{max} = 1.$ Our goal in this work is to determine the noisy capacity region of a switch that applies purification with $\alpha_{max} \ge 1.$ In this Section, we highlight the differences between the assumptions and proofs in our work to that of \cite{Vasantam22} and refer the interested readers to \cite{Vasantam22} for a more elaborate study. We use the same notations as used in \cite{Vasantam22} for simplicity and ease of understanding. We summarize notations in Table \ref{table:smstab}.  

\subsection{Proof of Theorem \ref{ps:capacity} and Theorem \ref{ps:mw}}
The proofs for the PS architecture are very similar to that obtained in \cite{Vasantam22} for $\alpha_{max} = 1$. When $\alpha_{max} = 1$ and there is no notion of noise, the random variable $\pmb{\tilde{T}}$ is Bernoulli distributed. When $\alpha_{max} > 1,$ and purification is applied, $\pmb{\tilde{T}}$ follows the distribution as mentioned in Equation \eqref{eq:tilde}. Following the capacity calculation and stability proof of max-weight protocol in  \cite{Vasantam22} with the distribution of $\pmb{\tilde{T}}$ as expressed in \eqref{eq:tilde} yields Theorem \ref{ps:capacity} and \ref{ps:mw}.

We now outline the proofs for the SP architecture.

\subsection{Proof of Theorem \ref{sp:capacity}}
We assume that link level entanglements decohere after one time slot. Hence, only $S_{ij}(n) = \min\{W_{ij}(n), Q_{ij}(n)\}$ number of entanglement swaps are performed in time slot $n$. Thus, $B_{ij}(n) \le S_{ij}(n)$. Also, $B_{ij}(n) = \sum_{k=0}^{S_{ij}(n)} Z_k,$ where $\Pr[Z_k = 1] = q_{ij}$ and $\Pr[Z_k = 0] = 1 - q_{ij}.$ Let $D_{ij}(n) = G(B_{ij}(n))$ where the function $G(\cdot)$ captures the effect of purification. $Q_{ij}(n)$ evolves as $Q_{ij}(n+1) = Q_{ij}(n) - D_{ij}(n) + A_{ij}(n)$. We start with the assumption that the process $\{\pmb{Q(n)}\}$ is an irreducible Markov chain. The conditions that a scheduling policy should satisfy for this assumption can be found in \cite{Vasantam22, Tassiulas97}. For the process $\{\pmb{Q(n)}\}$ to be stable, the following condition should hold true.
\begin{align}
    &\pmb{\lambda} \le \mathbb{E}[\pmb{D(n)}] = \mathbb{E}[\mathbb{E}[\pmb{D(n)}|\pmb{Q(n)}, \pmb{T(n)}]]\nonumber \\
    &= \sum\limits_{\pmb{a}\in \mathcal{A}, \pmb{a}\neq 0}P_{\pmb{T(n)}}(\pmb{a}) \sum\limits_{\pmb{\tau}}P_{\pmb{Q(n)}}(\pmb{\tau})\nonumber\\ 
    &\quad\quad\quad\quad\quad\quad\quad\quad\quad\quad\cdot\mathbb{E}[\pmb{D(n)}|\pmb{Q(n)} = \pmb{\tau}, \pmb{T(n)} = \pmb{a}]\label{eq:main}
\end{align}
We now derive $\mathbb{E}[D_{ij}(n)|\pmb{Q(n)}, \pmb{T(n)}]$ for all $(i, j)$ as follows.
\begin{align}
    &\mathbb{E}[D_{ij}(n)|\pmb{Q(n)}, \pmb{T(n)}] = \mathbb{E}[G(B_{ij}(n))|\pmb{Q(n)}, \pmb{T(n)}]\nonumber \\
    & = \mathbb{E}[\mathbb{E}[G(B_{ij}(n))|B_{ij}(n), \pmb{Q(n)}, \pmb{T(n)}]],
\end{align}
\noindent and $\mathbb{E}[G(B_{ij}(n))|B_{ij}(n), \pmb{Q(n)}, \pmb{T(n)}] = \mathbb{E}[Y|X = B_{ij}(n)].$ Therefore,
\begin{align}
    &\mathbb{E}[D_{ij}(n)|\pmb{Q(n)}, \pmb{T(n)}] = \mathbb{E}[\mathbb{E}[Y|X = B_{ij}(n)]| \pmb{Q(n)}, \pmb{T(n)}],\nonumber\\
    &=\sum\limits_{k=0}^{S_{ij}(n)} \mathbb{E}[Y|X =k] {S_{ij}(n) \choose k}q_{ij}^k(1-q_{ij})^{S_{ij}(n) - k},\nonumber\\
    &\le \sum\limits_{k=0}^{W_{ij}(n)} \mathbb{E}[Y|X =k] {W_{ij}(n) \choose k}q_{ij}^k(1-q_{ij})^{W_{ij}(n) - k}\label{eq:int-exp}
\end{align}
Substituting \eqref{eq:int-exp} in Equation \eqref{eq:main} and removing the conditioning on $\pmb{Q(n)}$ in Equation \eqref{eq:main} as discussed in \cite{Vasantam22}, we get the capacity region derived in Theorem \ref{sp:capacity}.
\subsection{Proof of Theorem \ref{sp:mw}}
To prove this, we apply Lyapunov stability of Markov chains using the Lyapunov function $V(\pmb{Q}(n)) = \sum_{ij}Q_{ij}(n)^2.$ We want to show that $\mathbb{E}[V(\pmb{Q(n+1)}) - V(\pmb{Q(n)})|\pmb{Q(n)}] \le -\epsilon||\pmb{Q(n)}||$. Thus we have
\begin{align}
    &\mathbb{E}[V(\pmb{Q(n+1)}) - V(\pmb{Q(n)})|\pmb{Q(n)}, \pmb{T(n)}] \nonumber \\
    &= \mathbb{E}[\pmb{Q(n+1)}\cdot\pmb{Q(n+1)} - \pmb{Q(n)}\cdot\pmb{Q(n)}|\pmb{Q(n)}, \pmb{T(n)}]\nonumber\\
    &= \mathbb{E}[(\pmb{Q(n+1)} - \pmb{Q(n)})\cdot(\pmb{Q(n+1)} - \pmb{Q(n)})\nonumber \\
    &+ 2(\pmb{Q(n+1)} - \pmb{Q(n)})\cdot\pmb{Q(n)}|\pmb{Q(n)}, \pmb{T(n)}]
\end{align}
It is easy to show that, $\mathbb{E}[(\pmb{Q(n+1)} - \pmb{Q(n)})\cdot(\pmb{Q(n+1)} - \pmb{Q(n)})|\pmb{Q(n)}, \pmb{T(n)}] \le d$, for some constant $d.$ Now we simplify, 
\begin{align}
&\mathbb{E}[(\pmb{Q(n+1)} - \pmb{Q(n)})\cdot\pmb{Q(n)}|\pmb{Q(n)}, \pmb{T(n)}]\nonumber\\
&=\mathbb{E}[(\pmb{A(n)} - \pmb{D(n)})\cdot\pmb{Q(n)}|\pmb{Q(n)}, \pmb{T(n)}]\nonumber\\
&=\mathbb{E}[\pmb{A(n)}\cdot\pmb{Q(n)}|\pmb{Q(n)}, \pmb{T(n)}] - \mathbb{E}[\pmb{D(n)}\cdot\pmb{Q(n)}|\pmb{Q(n)}, \pmb{T(n)}]\nonumber\\
&=\pmb{\lambda}\cdot\pmb{Q(n)} - \mathbb{E}[\pmb{D(n)}\cdot\pmb{Q(n)}|\pmb{Q(n)}, \pmb{T(n)}]\nonumber\\
&=\pmb{\lambda}\cdot\pmb{Q(n)} - \sum\limits_{ij}\mathbb{E}[D_{ij}(n)|\pmb{Q(n)}, \pmb{T(n)}]Q_{ij}(n)\label{eq:lyapunov}
\end{align}
\noindent Substituting \eqref{eq:int-exp} in \eqref{eq:lyapunov} and proceeding in a similar manner as in \cite{Vasantam22} establishes the Lyapunov stability condition.

\bibliographystyle{abbrv}
\bibliography{refs} 
\end{document}